\documentclass[final,9pt]{iiisconf}
\usepackage{multicol}
\usepackage{epsfig}
\usepackage{amsmath, amsthm, amssymb}
\newtheorem{theorem}{Theorem}[section]

\begin{document}
\title{A Game Theoretical Approach to Modeling Information Dissemination in Social Networks}
\author{%
\bf Dmitry Zinoviev, Vy Duong, Honggang Zhang\\%
\bf Mathematics and Computer Science Department, Suffolk University\\%
\bf Boston, Massachusetts 02114, USA%
}
\date{}
\maketitle

\section*{ABSTRACT}
One major function of social networks (e.g., massive online social networks) is the dissemination of information such as scientific knowledge, news, and rumors. Information can be propagated by the users of the network via natural connections in written, oral or electronic form. The information passing from a sender to a receiver intrinsically involves both of them considering their self-perceived knowledge, reputation, and popularity, which further determine their decisions of whether or not to forward the information and whether or not to provide feedback. To understand such human aspects of the information dissemination, we propose a game theoretical model of the information forwarding and feedback mechanisms in a social network that take into account the personalities of the sender and the receiver (including their perceived knowledgeability, reputation, and desire for popularity) and the global characteristics of the network.

\vskip\baselineskip\noindent
{\bf Keywords:} Social Network, Learning, Knowledge, Game Theory, Popularity, Reputation, Rumor.

\section{INTRODUCTION}
A social network is an ensemble of communicating personalities based on the concept of social proximity.  The members of a social network can form communities~\cite{spertus2005}, influence other members~\cite{crandall2008}, and engage in other social activities. 
One major function of social networks (in particular, massive online social networks) is the dissemination of information such as scientific knowledge, news, and rumors~\cite{1099206,nekovee2008,watts2002,zanette2002}. As an important form of social organization, information can shape public opinion, inform and misinform the society, cause panic in a society, promote products, etc.~\cite{nekovee2008}.
Information can be propagated by the members of the network via natural connections in written, oral or electronic form. 

Due to its importance, information dissemination or diffusion has been one of the focuses in social network research. For example, theories of rumor spreading are proposed in ~\cite{nekovee2008, zanette2002} to study the information dissemination.
Game theoretical approach to information propagation (namely, to learning) has been suggested by Gale {\it et al.}, Acemoglu {\it et al.}~\cite{NBERw14040,gale2003}. Ellwardt and van Duijn explored gossiping in small (organizational) social networks~\cite{ellwardt2009}.
Since information dissemination (and other various social network activities)
are supported by the structural organization of social networks, social network topology receives a lot of research attention. For example, 
an effect of network topology on the information diffusion was observed in~\cite{hirshman2009}: sparse networks are more effective for information entrance, and clustered (cellular) network structure decreases information diffusion.

In this paper, we propose a game theoretical model for the information passing from a sender to a receiver in an online social network. The novelty of our model is that psychological characteristics are explicitly modeled in information dissemination. 
In our model, information passing intrinsically involves both parties considering their psychological characteristics: self-perceived knowledge, reputation, and popularity---which further determine their decisions of whether or not to forward the information and whether or not to provide feedback. The decisions are also based on the global properties of the network, such as the overall quality of information and the way unreliable rumors are treated by the network members.

Our work differs from ~\cite{nekovee2008,zanette2002}, where a simple ``infection'' model was proposed for rumor spreading, and analytical and numerical solutions have been discussed. In particular, \cite{zanette2002} shows that the fraction of network population that never learns certain news, is a function of the fraction of ``refractory,'' or disinterested network members. These works identify three groups of network members: ignorants, spreaders, and stiflers (or susceptible, infected, and refractory). When a spreader contacts another spreader or a stifler, the initiating spreader becomes a stifler at a rate $\alpha$. When a spreader (sender) contacts an ignorant (receiver), the ignorant becomes a spreader at a rate $\beta$~\cite{nekovee2008}. The values of the parameters $\alpha$ and $\beta$ are assumed to be taken from the outside of the model.

Feedback in information dissemination is explicitly considered as strategic moves in our game theoretical model. Related to our work, Lampe {\it et al.} also analyzed the mechanism of feedback, its influence on the members of online communities, and its role in learning transfer~\cite{1099206}. Similar concept of social influence, but in the context of community building, has been researched by Crandall {\it et al.}~\cite{crandall2008}.

\section{\label{model}MODEL OF INFORMATION DISSEMINATION}

In this section, we present our model for characterizing users in social networks and for the information dissemination between a sender and a receiver. Each user has a utility function which is a combination of knowledge, reputation, and popularity. The information passing between a pair of users involves learning, feedback, and utility updating of both the sender and the receiver. Based on the model introduced in this section, we propose a game theoretic model to study the strategic actions of users in Section~\ref{game_model} and then further look at the information dissemination in a network setting involving a large number of users in Section~\ref{numerical}.

\subsection*{Network}
A social network can be modeled as a directed or undirected graph $G=(Z,Y)$, where each vertex in the vertex set $Z$ represents a human user or an actor (denoted as $A$), and each edge in the edge set $Y$ represents a relationship between a pair of actors. In online social networks, the relationship is usually called ``friendship''~\cite{zinoviev2009}. For most offline and many online social networks (e.g., MySpace, Facebook, LinkedIn), $G$ is undirected. In this paper, we assume that the network is connected and undirected.

\begin{figure}[b!]\centering
\epsfig{file=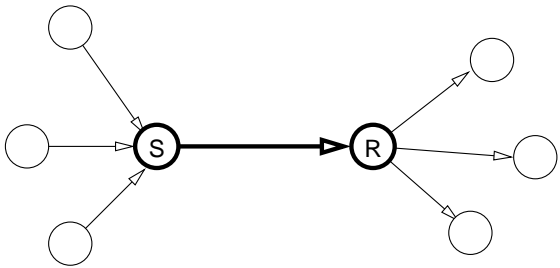,width=0.9\columnwidth}
\caption{\label{network}Communication between sender $S$ and receiver $R$.}
\end{figure}

We consider a homogeneous network where all actors or vertices are equally important, i.e., the identity of a node is not important. This allows us to consider the proliferation of information from an arbitrary
actor $S$ (sender) to another arbitrary actor $R$ (receiver), as long as the sender and the receiver are connected in the network. Figure~\ref{network} shows the information propagation in an online social network.

\subsection*{Assertion and Knowledge}

We suppose that there is a discrete finite set $\boldsymbol{F}$ of $N$ assertions that is to be shared among a network of users or actors. We assume that all assertions are equally important.

An assertion intrinsically can be true or false. We use $\varphi$ to denote the probability that a randomly selected assertion from the assertion set is true. Note that $\varphi$ is a system-wide parameter. The value of $\varphi$ is higher for formal communities (e.g., a scientific community) and lower for informal communities (e.g., a chatroom).

In addition to its intrinsic objective truthfulness, each assertion known to an actor has a subjective belief associated with that actor: the actor either believes that the assertion is true or false, or fails to make a decision for herself. An assertion that an actor does not know to be true or false is essentially a rumor---a ``story\ldots without any known authority for [its] truth''~\cite{webster1913}.

An actor $A_i\in Z$ knows $F_i\le N$ assertions, among which $A_i$ subjectively (privately) believes $F_i^+$ assertions are true and $F_i^-$ assertions are false, and treats all other known assertions $F_i^\circ$ as rumors (obviously, $F_i^++F_i^\circ+F_i^-\equiv F_i$). The probability of a randomly chosen assertion $\Phi\in\boldsymbol{F}$ (from the whole assertion set for the network) being known by $A_i$ is given by $f_i=F_i/N$. Similarly, we can find the probability of an assertion believed as true, rumor, or false as $f_i^+=F_i^+/F_i$, $ f_i^\circ=F_i^\circ/F_i$, and $f_i^-= F_i^-/F_i$, respectively.

Based on its own known set of assertions $\boldsymbol{F}_i$, actor $A_i$ forms its self-perceived \emph{knowledge} $K_i$ as follows:
\begin{equation}
 K_i=F_i^++F_i^-+\lambda F_i^\circ
\end{equation}
Here, $\lambda\in[0,1]$ is a ``rumor discount'' coefficient to capture the extent that an actor is willing to treat rumors as part of her knowledge. An actor with $\lambda=0$ completely discards rumors from her self-perceived knowledge, whereas an actor with $\lambda=1$ treats all her known rumors as if they were trustworthy assertions. Similar to $F_i$, knowledge $K_i$ is also bounded by $N$, and $F_i\le K_i\le N_i$.
In this paper, we assume that the value of $\lambda$ is the same for all actors in a social network.
We further normalize an actor's knowledge as $k_i=K_i/N$.

It is important to understand that the truthfulness of an assertion is not necessarily in agreement with the actor's beliefs. A good example of the disagreement would be the concept of the Flat World that was intrinsically false, but perceived as true by the vast majority of actors before the Enlightenment.

Based on their perceived knowledge, we roughly classify all actors into ``Ignoramuses'' (low $K_i$), ``Mediocres'' (medium-ranged $K_i$), and ``Gurus'' (high $K_i$). No actor in the network can definitely tell whether an assertion is true or false. However, we assume that there exists an external verification mechanism (an ``oracle'') that knows the definite answer.

\subsection*{Utility and Personality}

The information passing from a sender to a receiver intrinsically involves both of them considering their \emph{self-perceived knowledge, reputation, and popularity}, which further determines their decisions of whether or not to forward the information and whether or not to provide feedback. An actor's self-perceived knowledge, reputation, and popularity collectively form its \emph{utility}. The weights an actor put on these three utility components characterize this actor's \emph{personality}. We will give a detailed explanation of these concepts in this section.

We use a non-negative real number to measure an actor $A_i$'s \emph{reputation} $C_i$---``overall quality or character as seen or judged by people in general''~\cite{webster1913}.
Lower values of $C_i$ mean lower trustworthiness in $A_i$, suggesting that opinions expressed by $A_i$---such as information and feedback---were questionable in the past and should be considered with a grain of salt. In the extreme case of $C_i=0$, actor $A_i$'s beliefs are completely not credible. On the contrary, higher values of $C_i$ mean that $A_i$'s subjective evaluation of assertions has been regarded as historically highly credible.

We measure an actor's \emph{popularity} $P_i$ using another non-negative real number.
An actor's popularity in a social network is somewhat synonymous to social visibility~\cite{parkhurst98}: $P_i=0$ corresponds to an actor who does not speak in public
and in general is not even known to exist. An actor with high popularity enjoys popular attention. We assume that there is no correlation between $P_i$ and $C_i$ for the same actor $A_i$ and that popularity is time sensitive in the sense that it decays or discounts over time.
We suppose that an actor's popularity decays by $\Delta P=-\delta$  per unit time unless the actor participates in information exchange with other nodes, such as sending out some assertion as a sender or sending feedback as a receiver.

We further define normalized reputation and normalized popularity as $c_i=C_i/N$ and $p_i=P_i/N$. Note, in our study, we choose $N$ to be sufficiently large such that both $c_i$ and $p_i$ are no larger than one.

We believe that the purpose of a rational actor being a member of a social network is to maximize her utility $U_i$, defined as a convex combination of knowledge, reputation, and popularity with coefficients $0\le\kappa,\sigma,\pi\le1,\kappa+\sigma+\pi=1$: 
\begin{equation}
 U_i=\kappa K_i+\sigma C_i+\pi P_i.\label{payoff}
\end{equation}

We use a particular set of coefficients $\{\kappa,\sigma,\pi\}$
to characterize a particular type of actors' personality. For example, $\kappa=\sigma=0,\pi=1$ describe a network of ``Internet trolls'' (actors, for whom bloated popularity is the primary goal of networking). On the other hand, $\kappa=\sigma=0.5,\pi=0$ probably corresponds to a scientific community of knowledge seeking altruists who care about their reputation and wisdom, but not about being quoted or even published.

In this paper, we focus on a homogeneous network where all actors' have the same utility function. We understand that in a real social network, actors are most probably heterogeneous. We leave the heterogeneous network as future work.


\subsection*{Information transmission between a pair of actors}

During the passing of an assertion from an actor (the sender) to another actor (the receiver), both actors update their self-perceived knowledge, reputation, and popularity. This process involves evaluating knowledge, learning assertions, sending feedback, and updating reputation and popularity. We now describe in this section the basic steps that are performed in this process, but leave the discussion on the actors' the strategic decision making in the next section.

We assume that initially each actor pre-learns a random collection of assertions, which she randomly classifies as true assertions, false assertions or rumors, and that a communication or assertion passing between a sender and a receiver always takes a unit time.

\subsubsection*{Evaluating Knowledge}

When an actor $S$ has an assertion $\Phi\in \boldsymbol{F}_S$ to share with her neighbors, she will decide whether to forward $\Phi$ to a neighbor or hold it indefinitely\footnote{Other strategies, such as holding the assertion just for a limited time and then forwarding it, will be studied in future.}, in order to maximize her utility, as defined by~(\ref{payoff}). If $S$ sends $\Phi$ to her neighbor $R$ (receiver), then $R$ may choose to respond to $S$ with either positive or negative feedback $\Psi$.

Upon receiving $\Phi$ from the sender, the receiver $R$ may or may not be able to decide if $\Phi$ is true or false, based on the following considerations:
\begin{itemize}
\item The receiver's self-perceived knowledge.
\item The probability of $\Phi$ being true by nature (an intrinsic or objective characteristic but not a subjective belief by any actor), given by the system-wide parameter $\varphi$.
\item The sender's reputation, $C_S$.
\item The sender's opinion on $\Phi$: for an arbitrary assertion, the probability of that assertion being perceived by the sender as true or false is given by $f_S^+$ or $f_S^-$; the sender has no definite opinion on the assertion with the probability of $f_S^\circ$.
\end{itemize}

Let $g^+$ and $g^-$ be the probabilities of $R$ deciding that $\Phi$ is true or false respectively, and $g^\circ$ be the probability of $R$ failing to decide on $\Phi$ (i.e. declare it as a rumor). Note that $g^++g^-+g^\circ=1$, because eventually $R$ has to make some decision. If both the sender $S$ and the receiver $R$ have complete knowledge of all assertions (i.e., $k_R=k_S=1$), then we have:
\begin{equation}
\begin{split}
g^+&=f_S^+=f_R^+=\varphi,\\
g^-&=f_S^-=f_R^-=1-\varphi,\\
g^\circ&=f_S^\circ=f_R^\circ=0.\label{ideal}
\end{split}
\end{equation}
If $R$ is an Ignoramus ($k_R=0$), we assume that $R$ chooses to trust the sender's opinion $f_S^{\{+,-,\circ\}}$, discounted by the sender's reputation $c_S$. That is,
\begin{equation}
\begin{split}
g^+&=c_S f_S^+,\\
g^-&=c_S f_S^-,\\
g^\circ&= 1-g^+-g^-.
\end{split}
\end{equation}
In all other cases where ($0<k_R<1$), we assume that  $g^{\{+,-\}}$ are weighted sums defined as:
\begin{equation}
\begin{split}
g^+&=k_R\varphi+\left(1-k_R\right)c_S f_S^+,\\
g^\circ&=\left(1-k_R\right)\left(c_S f_S^\circ+\left(1-c_S\right)\right),\\
g^-&=k_R\left(1-\varphi\right)+\left(1-k_R\right)c_S f_S^-,\label{g-def}
\end{split}
\end{equation}
where the receiver's knowledge $k_R$ is a weighting factor.

Note, in the above probability computation, we can guarantee that $g^+,g^-, g^\circ$ are no larger than 1 as $r_s$ is no larger than 1 (as mentioned before, we choose $N$ to be sufficiently large).

\subsubsection*{Learning Assertion and Updating Knowledge and Popularity}

Once an assertion is transmitted from $S$ to $R$, both $S$ and $R$'s knowledge and popularity may change due to the transmission: $S$ informs $R$ of a potentially new assertion $\Phi$, and $R$ corrects $S$'s opinion on $\Phi$. When $R$ receives an assertion $\Phi$, one of the following three scenarios can happen:
\begin{enumerate}
 \item $R$ knows about $\Phi$, and his new opinion on the assertion, $g^{\{+,-,\circ\}}$, matches his existing opinion $f^{\{+,-,\circ\}}$. In this case, $R$ is not interested in $\Phi$. He discards the assertion and does not improve $S$'s popularity. The probability of this scenario is $p_1=f_R\left(g^-f_R^-+g^+f_R^++g^\circ f_R^\circ\right)$, where $f_R$ is the total number of assertions known to $R$.
\item $R$ knows about $\Phi$, but will reconsider his opinion (``I was told the world is flat; hey, I thought it was a joke, but the guy who told me the news traveled a lot! Perhaps he's right!''). In this scenario, $R$ re-labels $\Phi$ in the assertion set $\boldsymbol{F}_R$ and gives $S$ a popularity increase of $1$. Re-labeling $\Phi$ does not change $F_R$, but it can change $K_R$ (when $\Phi$ becomes a rumor or ceases to be a rumor). The probability of this event is $p_2=f_R\left(1-\left(g^-f_R^-+g^+f_R^++g^\circ f_R^\circ\right)\right)$.
\item Finally, $\Phi$ can be completely new to $R$. Then $R$ stores $\Phi$ in the assertion set $\boldsymbol{F}_R$ and gives $S$ a popularity increase of $1$. The number of assertions of $R$'s increases, and the amount of knowledge of $R$'s increases, too. This scenario happens with  probability $p_3=1-f_R$.
\end{enumerate}
Then, we see that the overall knowledge change at $R$'s side is given by:
\begin{equation}
\begin{split}
 \Delta K_R=&\lambda\left(1-f_R\right)\\
+&\left(1-\lambda\right)\left(g^-+g^+-f_R\left(f_R^++f_R^-\right)\right).\label{delta-KR}
\end{split}
\end{equation}

Notice that if $S$ knows {\em a priori} that $R$ is already in the possession of the assertion that $S$ is about to share with $R$, then the probability of the above third scenario is $0$, and~(\ref{delta-KR}) changes accordingly:
\begin{equation}
\begin{split}
 \Delta K_R'=\left(1-\lambda\right)\left(g^-+g^+-f_R\left(f_R^++f_R^-\right)\right).\label{delta-KR1}
\end{split}
\end{equation}
This situation may arise when $S$ has reinterpreted the assertion. We will not use~(\ref{delta-KR1}) in our further analysis.

Unlike the receiver $R$, the sender $S$ updates her knowledge based on the feedback provided by $R$ (if any). The number of assertions at the sender's side does not change, only $S$'s opinion on them can change and $S$'s perceived knowledge that depends on the opinion. If $S$ trusts $R$ (because of $R$'s high reputation $C_R$), then $S$ can change her belief about $\Phi$. The change is given by the formula:
\begin{equation}
\begin{split}
 \Delta K_S=c_R\left(1-\lambda\right)\left(g^-+g^+-\left(f_S^++f_S^-\right)\right).
\end{split}
\end{equation}

The total popularity premium to sender $S$ is given by:
\begin{equation}
 \Delta P_S=1-f_R\left(g^+f_R^++g^-f_R^-+g^\circ f_R^\circ\right).\label{delta-P}
\end{equation}
The receiver $R$ enjoys the popularity growth of $1$ only if $R$ chooses to send feedback to $S$ (e.g., by commenting her original assertion).

\subsubsection*{Sending Feedback and Updating Reputation}

A feedback mechanism is used by $R$ to affect the reputation of the sender $S$ and, eventually, $R$'s own reputation.

We use an actor $A_i$'s reputation $C_i$ as the measure of $A_i$'s ability to inspire belief. Actor $A_i$'s successful prediction of the true nature of an assertion should increase $C_i$, while an incorrect prediction should decrease it. Unfortunately, no actor is the network knows the true value of a random assertion (even a Guru can predict that an assertion is true only with the probability of $\varphi$). That is why we need an external oracle that compares $R$'s perception of assertion $\Phi$ with the true nature of that assertion and concludes if the evaluation was successful or not. If $R$ is a Guru (high $k_R$), he makes the right choice and earns a reputation value $1$. If $R$ is an Ignoramus, the best he can do is to trust $S$ (to the extent of the sender's reputation). In the latter case, $R$ gets a reputation boost of $\left(1\times c_S\right)$ if $S$'s belief of $\Phi$ is accurate (which happens with the probability of $\varphi f_S^++(1-\varphi)f_S^-$), and a penalty of $\left(-1\times c_S\right)$ otherwise (with the probability of $\varphi f_S^-+(1-\varphi)f_S^+$).  After an assertion is passed between the sender and the receiver, the expected change of the receiver's reputation is:
\begin{equation}
\Delta C_R=k_R+\left(1-k_R\right)\left(c_S\left(2\varphi-1\right)\left(f_S^+-f_S^-\right)\right).
\end{equation}

In the mean time, $R$ can affect $S$'s reputation by providing feedback in the following way: if $S$ and $R$ agree on their perceived nature of $\Phi$ (the true nature of $\Phi$ is not involved) and the receiver is a credible authority himself, then the sender's reputation improves by value $1$; otherwise, it decreases by value $1$. In other words:
\begin{equation}
\begin{split}
\Delta C_S=C_R\Big(&\left(1-2g^+-2g^-\right)\left(1-2f_S^+-2f_S^-\right)\\
&-2\left(f_S^+g^++f_S^-g^-\right)\Big).\label{delta-CS}
\end{split}
\end{equation}

It is quite possible that the change of knowledge $K$, numbers of known, true, false, and neutral assertions $F$, $F^+$, $F^-$, $F^\circ$, popularity $P$ or reputation $C$, resulting from a communication, will make one or more of these values greater than $N$ or less than 0. Our model is not designed to handle these situations and should be used only when each of these numbers is greater than 1 and less than $N-1$. This anomaly can be avoided by choosing $N$ to be large.

\section{TWO-PLAYER INFORMATION DISSEMINATION GAME}\label{game_model}

In the previous section, we gave a detailed analysis of the basic steps involved in the information transmission between two actors (a sender and a receiver) and described how the two actors update their utilities (including self-perceived knowledge, reputation, and popularity) depending on whether or not the sender transmits an assertion to the receiver and whether or not the receiver sends feedback to the sender. However, we have not answered this question yet: under what circumstances is the sender willing to transmit an assertion and when does the receiver decide to send feedback? In this section, we will address this question under the assumption that both actors
know that each of them attempts to maximize its own utility, and they are fully aware of the impact on their own utilities from any combination of their individual choices. 

Such a strategic interaction between the sender and receiver can be naturally modeled as a game with both actors being players.
More specifically, both actors play a rectangular $2\times 2$ game with a non-zero sum.
The utility change of both players in the game are given by the following payoff matrix:
\begin{equation}
 M=\begin{Vmatrix}
	\Delta U_S^{00},\Delta U_R&\vdots&\Delta U_S^{01},\kappa\Delta K_R-\pi\delta\\
\hdotsfor{3}\\
	-\pi\delta,-\pi\delta&\vdots&-\pi\delta,-\pi\delta\\
   \end{Vmatrix}\label{payoff_matrix}
\end{equation}
The rows correspond to the two available actions of the sender ($S_0$ and $S_1$), and the columns correspond to the two available actions of the receiver ($R_0$ and $R_1$).
The sender has to choose between two actions: to forward information ($S_0$) or not to forward information ($S_1$). The receiver also has to choose either to provide feedback ($R_0$) or not to provide feedback ($R_1$). The combination of $S_1$ and $R_0$ is not feasible as $R$ cannot provide feedback for a message that has not been sent by $S$.

It is not immediately clear if the two actors should play this game in a cooperative fashion or not. In an informal online environment (such as a chatroom, a Web forum, or a political discussion club), actors are not necessarily interested in optimizing their joint utility. Rather, their individual utilities are of the primary importance. Such a social behavior can be modeled as a non-cooperative game. On the contrary, in selected ``gated'' environments such as research seminars or hobbyist meetings, a receiver may want to avoid excessive negative feedback to not hurt the sender beyond necessity. In this situation, a cooperative game makes more sense. Nevertheless, in this work we are inclined to believe that a non-cooperative game better reflects the communication in the real world, at least as the first approximation.

By treating the information dissemination game defined in (\ref{payoff_matrix}) as a non-cooperative game, we have derived the following theorem to characterize the Nash equilibrium (if exists) of the game.
\begin{theorem}
 The game defined by the matrix $M$ has a pure strategy weak Nash equilibrium.
\begin{proof}
 We will prove the theorem by assuming that a mixed strategy exists for the sender and that strategy is a part of the Nash equilibrium. Let $x$ be the probability of S playing $S_0$ and $1-x$ the probability of playing $S_1$. The condition for the Nash equilibrium for the receiver is $x\left(\Delta U_R\right)+\left(1-x\right)\left(-\pi\delta\right)=x\left(\kappa\Delta K_R-\pi\delta\right)+\left(1-x\right)\left(-\pi\delta\right)$. From this follows that either $x=0$ or $\Delta U_R=\kappa\Delta K_R-\pi\delta$. In the former case, $S_1$ (``never forward'') is the pure optimal strategy. In the latter case, $x$ is irrelevant and can be chosen to be either 0 or 1 (in the numerical experiments, we select $x=1$: ``always forward'').

A similar proof can be constructed for the receiver.
\end{proof}
\end{theorem}

We use $\{S_i^*,R_j^*\}$ to denote the equilibrium strategy profile of
the game. At each such equilibrium, we can find each actor's state
variables such as their utilities, and normalized values popularity, reputation, and knowledge.

\section{\label{numerical}CASE STUDY: INFORMATION DISSEMINATION IN NETWORK}

We now explore the information dissemination process in a network with a large number of actors, based on the two-player game model introduced in the previous section. To this end, 
we conducted a number of experiments by simulating the information dissemination in discrete time steps on a complete bidirectional social network of $1000$ actors. At each time step, one actor $S$ is randomly chosen to play the two-player information dissemination game with another randomly chosen actor $R$. This is not an $n$-player game, but a series of independent 2-player games. The state values of both actors at Nash equilibrium for each game are recorded. Recall that we assume each such game can be completed in one time step. 

The model parameters of our simulations are as follows. The probability of an assertion being true is $\varphi=0.8$. The actor popularity decay factor is $\delta=0.1$. The rumor discount coefficient is $\lambda=0.5$. The maximum number of assertions in the network is $N=2000$.

In each experiment, the network was populated by actors of a certain type. In the first experiment, all actors were ``Internet trolls'': $\kappa=0.1$, $\sigma=0.1$, $\pi= 0.8$; in the second experiment, we modeled an ``Internet expert'' community: $\kappa=0.2$, $\sigma=0.7$, $\pi= 0.1$; finally, in the third experiment 50\% of the actors were randomly chosen to be ``experts,'' and the rest of them were ``trolls.'' In each experiment, a third of the actors initially started as ``Ignoramuses'' ($k=0.1$), another third as ``Mediocres'' ($k=0.5$), and the remaining third as ``Gurus'' ($k=0.9$). The initial values of reputation $r$, popularity $p$, and the fractions of true and false assertions $f^+, f^-$ were drawn uniformly at random between the minimum value of $0$ and the maximum values of $0.5$ for $f^-$ and $1$ for the other three parameters.

\begin{figure}[tb!]\centering
\epsfig{file=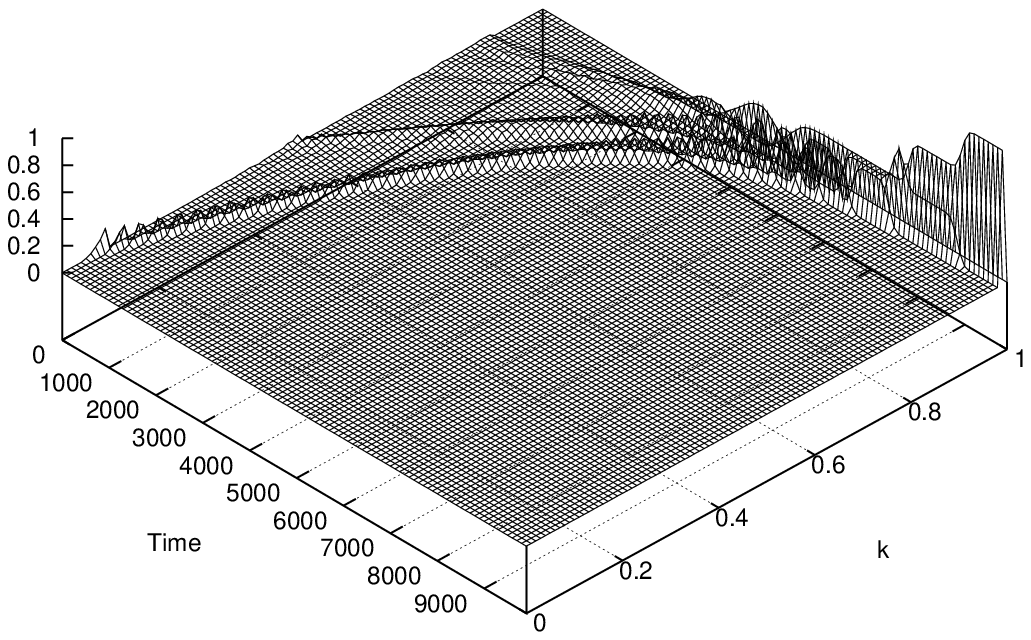,width=\columnwidth}
\vskip0.5\baselineskip
\epsfig{file=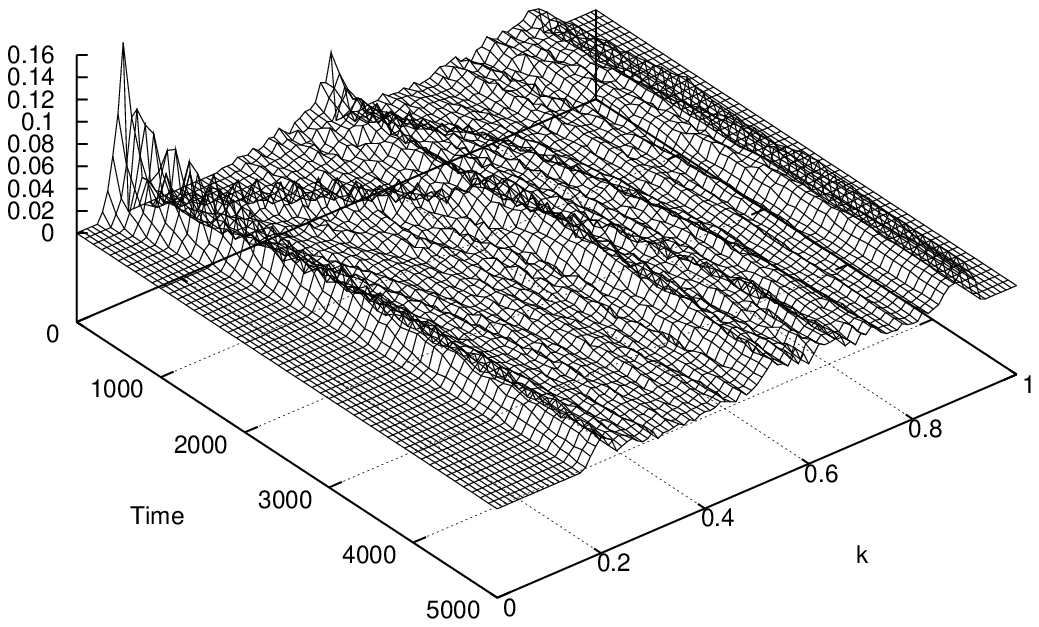,width=\columnwidth}
\vskip0.5\baselineskip
\caption{\label{knowledge-troll}Distribution of actors by $k$ as a function of simulation time. Top: ``troll'' community; bottom: ``expert'' community.}
\end{figure}

\begin{figure}[tb!]\centering
\epsfig{file=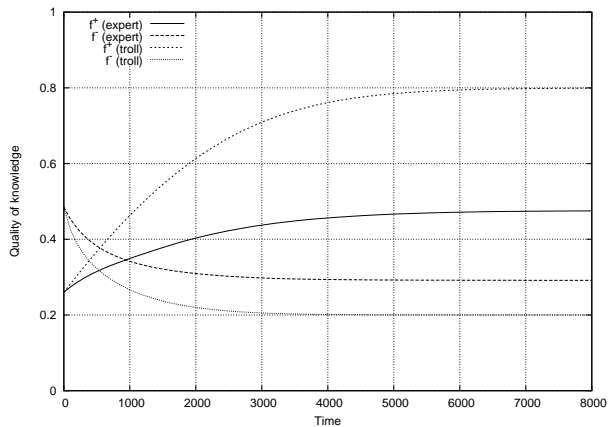,width=\columnwidth}
\caption{\label{quality}Quality of knowledge in the ``expert'' and ``troll'' communities as a function of time.}
\end{figure}

For the duration of 10 million assertion transmissions (on average 10,000 communications per actor), we monitored the distribution of $k$ in the network and the ``quality'' of knowledge---the average values of $f^+$ and $f^-$---as functions of the simulation time (the simulation time is defined as the average number of communications per actor). The results of  sample simulation runs for the first two experiments are shown in Figures~\ref{knowledge-troll} and \ref{quality}. The results of the third experiment (with the equal mix of ``expert'' and ``troll'' actors) are somewhere in between the two extreme cases, and they are not shown here.

One can tell from Fig.~\ref{knowledge-troll} that both models converge to stable states; however, the states are not the same. The ``troll'' community converges to the state of ``total knowledge,'' where all actors become fully knowledgeable ($f\approx k\approx 1$) after a finite number of iterations. The distribution of information in the ``expert'' community, on the contrary, changes (disperses) only marginally in time. The difference is due to the fact that the troll utility function mostly ignores reputation, but favors popularity, and trolls almost always choose to spread assertions without fearing the potentially harmful feedback. The expert utility function emphasizes reputation, thus preventing ``experts'' from distributing information of uncertain nature.

The quality of knowledge in the ``troll'' community eventually converges to the statistical levels determined by $f^+=\varphi=0.8$ and $f^-=1-\varphi=0.2$ in Fig.~\ref{quality}). This implies that using the knowledge evaluation mechanism as defined before, actors cannot classify assertions better than statistically possible. The expert network does not converge even to the statistically possible level for the reason described above. It does not follow from this observation that ``expert'' communities are slowing down or blocking information exchange: they are merely practicing a more cautious approach.

\begin{figure}[tb!]\centering
\epsfig{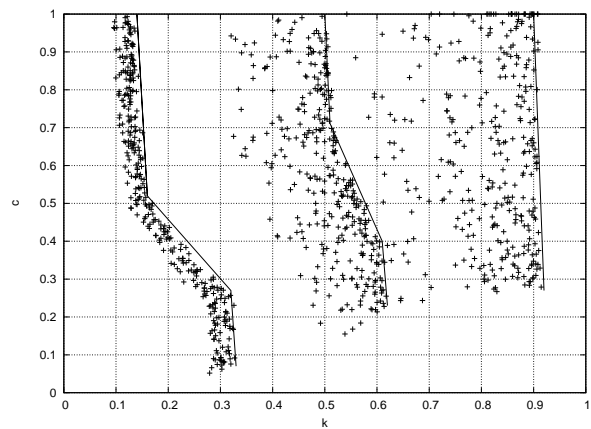}
\caption{\label{c-vs-k}Reputation and knowledge of ``expert'' actors at the simulation time of 800.}
\end{figure}

Another important difference between the models in Fig.~\ref{knowledge-troll} is the significant dispersion of the learning speeds in the ``expert'' community: a fair fraction of ``Ignoramuses'' and ``Mediocres'' learn much faster than the other actors in the same class. To understand the nature of this phenomenon, we recorded the personality parameters---namely, reputation $r$ and popularity $p$---of all actors at the time 800 (just after the bifurcation in the Fig.~\ref{knowledge-troll}). We did not see any correlation between $k$ and $p$. However, the correlation between $k$ and $r$ was overwhelming (Fig.~\ref{c-vs-k}).

The Figure suggests that actors with lower reputation learn faster, and the learning speed is lower for the actors with higher initial knowledge. This means that actors with initially low knowledge and low reputation actively gain knowledge to increase their utility, while those with initially high knowledge (``Gurus'') or high reputation have less incentive to learn new knowledge.

\section{\label{conclusion}CONCLUSION AND FUTURE WORK}

In this paper we proposed a game theoretical model of information dissemination in social networks. The model takes into account both personal traits of the users of the network (the desire for self-perceived knowledge, reputation, and popularity) and the properties of the disseminated information (in particular, its overall truthfulness). The feedback mechanism is used to control the reputation of information senders and deter them from disseminating unconfirmed rumors.
The model is mathematically represented by a $2\times 2$ non-zero sum, non-cooperative game with a pure strategy Nash equilibrium, where  the available actions to a sender is to forward an assertion or information or hold it indefinitely and a receiver needs to decide whether to provide feedback or not.
Our numerical experiments show that a massive social network of actors communicating using the proposed model demonstrates intuitively acceptable aggregate learning behavior.

To extend and generalize our model, we proposed the following future research directions. 
\begin{itemize}
 \item The variability of  $\kappa$, $\sigma$, and $\pi$  for different actors in a network will be considered in our future work. 
\item A common strategy in real social network is not to forward an assertion immediately or hold it indefinitely, but to hold for a limited time. We will incorporate the delayed strategies in our game theoretic model.
\end{itemize}

\section*{Acknowledgment}
This research has been supported in part by the College of Arts and Sciences, Suffolk University, through an undergraduate research assistantship grant.

\bibliographystyle{acm}
\bibliography{cs}
\end{document}